\documentclass[twocolumn,prd,noshowpacs]{revtex4}
\usepackage{amssymb}
\usepackage{amsmath}
\usepackage{amsfonts}
\usepackage{graphicx, float}

\begin{document}

\title{Reply to Comment on ``Inflation with a graceful exit and entrance driven by Hawking
radiation"}

\author{Sujoy K Modak}
\email{sujoy.kumar@correo.nucleares.unam.mx}
\affiliation{Instituto de Ciencias Nucleares, Universidad Nacional Aut\'{o}noma de M\'{e}xico,\\ Apartado Postal 70-543, Distrito Federal, 04510, M\'{e}xico.}
\author{Douglas Singleton}
\email{dougs@csufresno.edu}
\affiliation{Physics Department, CSU Fresno, Fresno, CA 93740 USA}

\date{\today}

\begin{abstract}
The comment \cite{comment} raises two points in regard to our work \cite{modak}. The first is that one cannot use the tunneling picture to obtain the temperature and particle production rate in the Friedman-Robertson-Walker (FRW) space-time. The second comment raised by \cite{comment} is that the Hawking-like radiation model for inflation 
presented in \cite{modak, modak2} is inconsistent with the observed scalar and tensor perturbation spectrum. We show that
the first comment is beside the point -- we do not use the tunneling method in our works \cite{modak, modak2}. 
The second criticism in \cite{comment} comes from the author evaluating quantities at different times -- he
evaluates the parameters of our model at the {\it beginning} of inflation and then compares this with the scalar and 
tensor perturbations evaluate at {\it the horizon exit} point.            
\end{abstract}

\maketitle

In \cite{comment} the author raises two points in connection with our work in \cite{modak} where we proposed 
a mechanism for inflation based on the Hawking-like radiation of Friedman-Robertson-Walker (FRW) space-time
(see also the earlier paper \cite{modak2} where this model of inflation is introduced). The 
first criticism of \cite{comment} is that one is not justified in using the tunneling/WKB approach 
introduced in \cite{wilczek, paddy} to calculate the temperature and particle creation rate of FRW space-time. 
In fact in our work \cite{modak} nor in our earlier paper on the subject \cite{modak2} did we use
the tunneling method to obtain the temperature or particle creation rate or anything else. The
Hawking-like temperature for FRW space-time which we use in \cite{modak, modak2} has been derived in 
by several researchers and using a host of different methods all which give the same temperature of 
\begin{equation}
\label{t-frw}
T_{FRW} \approx \frac{\hbar c}{2 \pi k_B r_A}
\end{equation}
where $r_A$ is the apparent horizon and the approximation is used since we are neglecting a time which
depends on th time variation of $r_A$ (the full expression for $T_{FRW}$ can be found in \cite{modak, modak2}).
There are several different derivations of $T_{FRW}$ which all comes give the same temperature: (i) in
\cite{cai} thermodynamic arguments are used to arrive at \eqref{t-frw}; (ii) in the works \cite{cai2, tao, tao2, hayward}
various forms of the tunneling method (null geodesic, Hamilton-Jacobi) are used; (iii) in \cite{yu} wave modes in an FRW
background are studied. All these works use different methods to arrive at the same temperature for the temperature
from an FRW space-time. Further in the the papers \cite{parker} and in the monographs \cite{BD,TP} quantum 
creation and annihilation operators are studied in FRW space-time and it is found that particle creation 
does occur in some special cases. Also, given that the surface gravity for an FRW space-time is given approximately by
$\kappa =1/r_A$ one can see that \eqref{t-frw} is in accord with the standard relationship between surface gravity
at a horizon and Hawking temperature namely
\begin{equation}
\label{t-gen}
T \approx \frac{\hbar c \kappa}{2 \pi k_B } ~.
\end{equation}
Finally, it is important to note that in the de-Sitter phase the notion of Hawking temperature 
is very well defined \cite{gh}. In our model it is sufficient to get inflationary behavior and then a 
graceful exit to the next phase from our effective de-Sitter phase. 
Therefore regardless of the issue of the existence of a well defined notion of temperature/particle 
creation in the subsequent phases, our model does work very well to explain the
effective de-Sitter inflationary phase. Indeed there are similar works on 
deflationary model which address the exit from the inflationary stage using Hawking temperature of the inflationary
de-Sitter phase \cite{lima}. Thus the first criticism of the comment \cite{comment} -- that we are not justified 
in using the tunneling method to obtain the Hawking-like temperature and particle creation rate of an FRW 
space-time -- is beside the point since in fact we do not use the tunneling method in \cite{modak, modak2}. 
As a side note we want to point out two issues related to the comment \cite{comment}: (i) The author incorrectly 
gives the particle creation/tunneling rate in the tunneling picture as
\begin{equation}
\label{gamma}
\Gamma = \exp \left[- \frac{2}{\hbar} Im \int p dr \right] 
\end{equation}
(see equation (1) of \cite{comment} where $\int p dr  = S$ ($p$ is the canonical momentum in the space-time). The
correct expression for the particle creation/tunneling rate is \cite{akhmedov}
\begin{equation}
\label{gamma-2}
\Gamma = \exp \left[- \frac{1}{\hbar} Im \oint p dr \right] ~,
\end{equation}
where the loop integral crosses the horizon out and back rather than \eqref{gamma} which is just twice 
either the outgoing or ingoing path. The expression \eqref{gamma-2} is canonically invariant and is thus a 
proper quantum observable, while the expression in \eqref{gamma} is not canonically invariant. Further in certain
situations the two expressions give different numerical results leading to the incorrect Hawking temperature 
\cite{akhmedov-2}. In addition the author of \cite{comment} neglects the temporal 
contribution to the particle creation/tunneling rate which is crucial to obtaining the correct result as 
shown in \cite{pilling, tao}. (ii) The author states that since the FRW horizon is an apparent horizon
rather than an event horizon, that Hawking radiation does not occur. This is not true. In general for a 
dynamic spacetime, satisfying standard energy conditions, Hawking radiation originates from the apparent 
horizon which is the outermost trapped surface. In fact there exist a vast literature on two dimensional 
black hole evaporation where Hawking radiation originates from the apparent horizon \cite{2d} if one 
considers back reaction.
  
There is a weak point or assumption in the inflation mechanism proposed in \cite{modak} --
we used the results for the temperature and  particle creation rate for FRW space-time in 
a time range when the expansion rate is too large to completely trust 
the results based on slow expansion approximation. This is in some sense the reverse problem with the evaporation of
black hole via Hawking radiation. In the case of black holes one can not trust the calculation of the Hawking temperature and
particle emission as the black hole mass shrinks to zero. However we clearly state this as an assumption several
times in the paper and in the absence of a more rigorous calculation which would take 
into account back reaction this seems to us a reasonable assumption as a starting point. 

The second point the author makes appears to be that the scale of our inflationary model is near the
Planck scale rather than being at the more standard view of the inflationary energy scale i.e.
Grand Unified Theory (GUT) scale. He also comments that our model appears to greatly overestimate the tensor
fluctuations in the CMB. First we point out that there are other proposals which postulate inflation
closer to the Planck scale as opposed to the GUT scale such as the work \cite{loop} which uses loop
quantum gravity to investigate cosmology near the Planck scale. In detail the author of \cite{comment}
shows that there is an apparent conflict between our form of inflation and the size of the scalar and
tensor fluctuations observed by COBE, WMAP, PLANCK \cite{bassett} \cite{ade}.  
Next the author uses our time dependent equation of state parameter, 
$\omega _c (t) = \frac{\hbar G^2}{45 c^7} \rho (t) \simeq \frac{4}{3}$, which is given in his equation (7), 
to obtain the relationship (in ``God-given natural units" where  $m_{Pl} =\sqrt{\frac{\hbar c}{G}}$ is 
the Planck mass \cite{comment})
\begin{equation}
\label{rho}
\frac{\rho (t)}{m_{Pl} ^4} \simeq 1 ~,
\end{equation}
where $\rho (t)$ is the time dependent energy density. The important point to note here is that this
is the value at the {\it beginning} of the inflationary era. The quantity in \eqref{rho} evolves with
time as \cite{modak}
\begin{equation}
\label{rho-2}
\rho (t) = \frac{D}{a^4 (t) + \frac{3 \alpha D}{4}} ~,
\end{equation}
where $D$ is a constant, $\alpha$ is connected with the particle creation rate and $a(t)$ is the
time dependent scale factor of the Universe, and we assumed that the early Universe is mostly 
radiation dominated -- setting the creation rate to zero ($\alpha = 0$) gives
$\rho (t) = D/a^4 (t)$ which is the energy density for a radiation dominated Universe. 
The form of $a(t)$ was determined in \cite{modak} and early on one had exponential expansion which 
transitioned to an ordinary radiation dominated expansion later on. The implicitly defined form of $a(t)$ 
is given in \cite{modak}. In integrating to solve for $a(t)$ there was an arbitrary integration constant
which allowed one to shift the starting time of inflation in our model (the length of the inflationary period 
was fixed by the creation rate $\alpha$, but we had the freedom to start the inflationary period 
at different times). 

The author of \cite{comment} points out an apparent conflict between \eqref{rho} and the scalar, ${\cal P}_R$,
and tensor, ${\cal P}_T$ perturbations defined as \cite{bassett}
\begin{equation}
\label{pr-pt}
{\cal P}_R \simeq \left( \frac{V^3}{m_{Pl} ^6 (V')^2} \right)_{k=a H}  ~~~;~~~
{\cal P}_T \simeq \left( \frac{V}{m_{Pl} ^4 } \right)_{k=a H} ~,
\end{equation}  
where $V (\phi)$ is the potential of the scalar inflaton field, $\phi$, and the primed means derivative with respect
to $\phi$. This quantities are evaluated at the {\it horizon exit} time i.e. the point when a given wave mode,
represented by $k$ crosses the Hubble radius. One can re-write the scalar perturbation spectrum ${\cal P}_T$ 
as \cite{comment}
\begin{equation}
\label{pt}
{\cal P}_T \simeq \left( \frac{\rho (t)}{m_{Pl} ^4} \right)_{k=a H} ~,
\end{equation} 
which is then the same as \eqref{rho} {\it except} that \eqref{pt} is evaluated
at the horizon exit whereas \eqref{rho} is at the start of the inflationary era . 
Now the scalar and tensor perturbations are related by
\begin{equation}
\label{r}
{\cal P}_T = r {\cal P}_R  
\end{equation}
From observations one finds \cite{bassett, ade} that $r<0.11$ and that ${\cal P}_R \approx 10^{-9}$ which by 
\eqref{r} yields ${\cal P}_T \le 10^{-10}$. Combining this observational result with \eqref{pt}
and our result \eqref{rho} gives an apparent conflict. However the comparison of \eqref{pt} with 
\eqref{rho} makes no sense at all since they are at completely {\it different times} -- \eqref{rho}
is at the outset of inflation while \eqref{pt} is evaluated at the later, horizon exit point. To see 
if there is a conflict one should evaluate the expression in \eqref{rho} also at the horizon exit
point $k = a H$. This would require going to the implicitly obtained expression for $a(t)$ given
in \cite{modak2}. Because of the arbitrary integration constant in $a(t)$ which allows one to shift 
the point at which inflation starts in our model one can always find parameters for which \eqref{rho}
{\it evaluated at the horizon exit point} agree with the observational requirement ${\cal P}_T \le 10^{-10}$.
One final point is that the definition of ${\cal P}_R$ and ${\cal P}_T$ in \eqref{pr-pt} presume that 
inflation is driven by a scalar inflaton field which some potential $V(\phi)$. In our model \cite{modak, modak2}
we have neither an scalar inflaton field nor any scalar potential so it is not clear that the analysis
starting with \eqref{pr-pt} can be directly applied to our model of inflation which is driven by 
particle creation. We are currently investigating this issue of the extend to which one can carry over the
results of inflation driven by scalar field to our model which is driven by particle creation.

{\bf Acknowledgment}: DS acknowledges useful discussion with Willard Mittleman.

\end{document}